\documentclass[12pt]{book}
\usepackage[dvips]{graphicx,color}
\usepackage{makeidx,tsukuba}
\makeauthorindex
\makeindex

\begin{document}

\BookTitle{\itshape The 28th International Cosmic Ray Conference}
\CopyRight{\copyright 2003 by Universal Academy Press, Inc.}
\pagenumbering{arabic}

\chapter{TeV Gamma-ray Observations and the Origin of Cosmic Rays: I}

\author{T.C.~Weekes\\ {Harvard-Smithsonian Center for Astrophysics,
Whipple Observatory,} \\ {P.O. Box 97, Amado, AZ 85645, USA}}

\section*{Abstract}
This is the first of three plenary talks with the same title
given at the 28th ICRC in Tsukuba, Japan in August, 2003.
A brief description of the  techniques for detecting 
gamma rays at TeV energies is followed by a summary of the 
observational status of the field. The expectations of the field 
from a cosmic ray perspective are compared with these early results. 
The majority of sources detected with some certainty are 
extragalactic; the observational status of these sources is summarized. 
The most complete set of observations are those dealing 
with the detection of blazars for which a catalog is presented.
This discipline is now established as a new branch of observational 
astronomy.
  
\section{Atmospheric Cherenkov Imaging Technique}

The remarkable advances in the detection of sources of TeV gamma
rays have come from the advances in ground-based detection
techniques. These advances are well
documented in the appropriate rapporteur papers of the last two
 decades of
ICRCs [2,4,8,12,14,23,24]. An account of
the early development of the technique has been given elsewhere
[26]. At this 28th ICRC in Tsukuba three speakers,
representing three of the major groups in the field, (Weekes for
the Whipple/VERITAS Collaboration, Kifune for the CANGAROO
Collaboration, Voelk for the HEGRA/HESS Collaboration) were asked
to speak on a common topic, the title of this paper. As might be
expected, the perspectives of the three speakers were quite
different. To minimize duplication, the speakers agreed that the
first speaker should concentrate on the extragalactic sky, the
second on the Galactic sources, and the third on an
interpretation of the observations.

\begin{figure}[ht!]
\centerline{%
\begin{tabular}{c@{\hspace{6pc}}c}
\includegraphics[width=5cm]{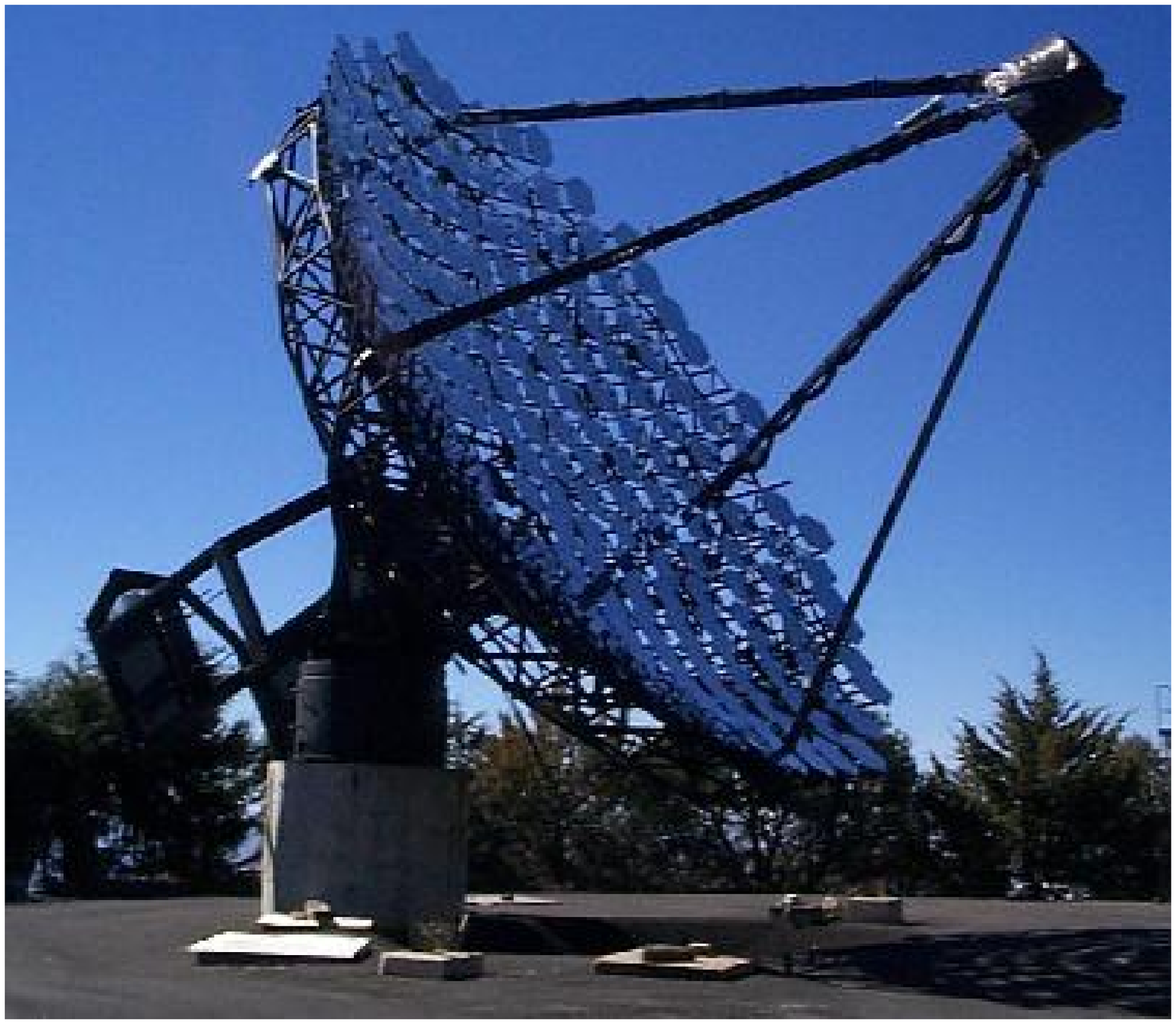} & 
\includegraphics[width=6cm]{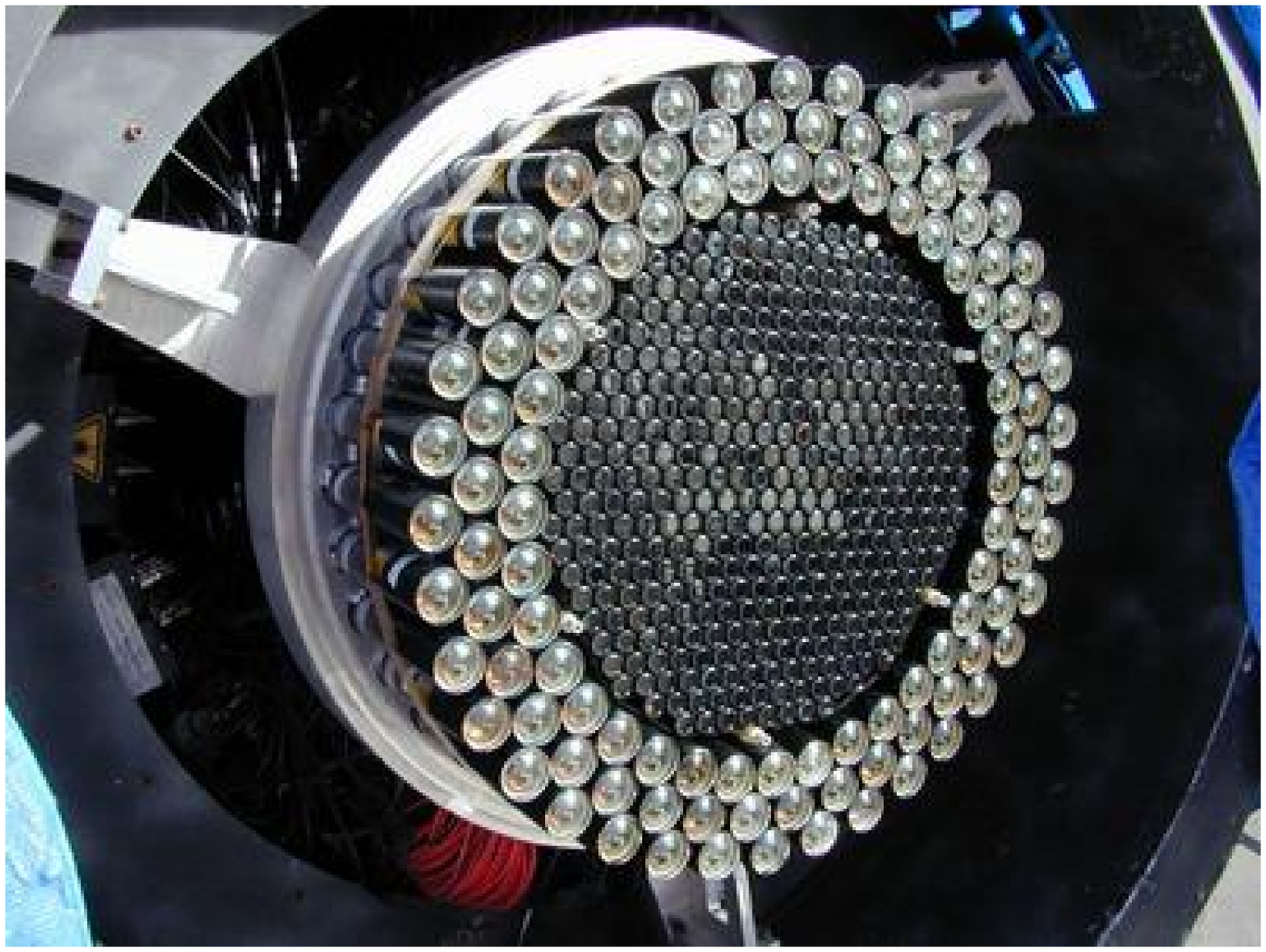} 
\end{tabular}
}
\caption{a. The Whipple 10m telescope which was built in 1968 
and is still in operation; b. The 490 pixel Whipple camera composed of 1.2 cm and
2.5 cm phototubes.}
\label{10m}
\end{figure}

Almost all of the TeV observational results discussed have come from
the use of the atmospheric Cherenkov imaging technique whereby 
the Cherenkov
light images from small air showers, as seen at ground level, are
recorded by fast cameras in the focal plane of large optical
reflectors. In practice the cameras are composed of arrays of
hundreds of small photomultipliers and the reflectors have apertures
ranging from 2 to 12 meters (Figure~\ref{10m}). 
At lower energies these observations
have been extended by observations with the so-called solar arrays
(CELESTE, STACEE, Solar-II)
and at higher energies by the particle detecting arrays (Milagro, Tibet, 
ARGO).  In the next decade further improvements in ground-based 
detection techniques are expected and these will dramatically change
the observational picture.

\begin{figure}[ht!]
\centerline{%
\begin{tabular}{c@{\hspace{6pc}}c}
\includegraphics[width=6cm]{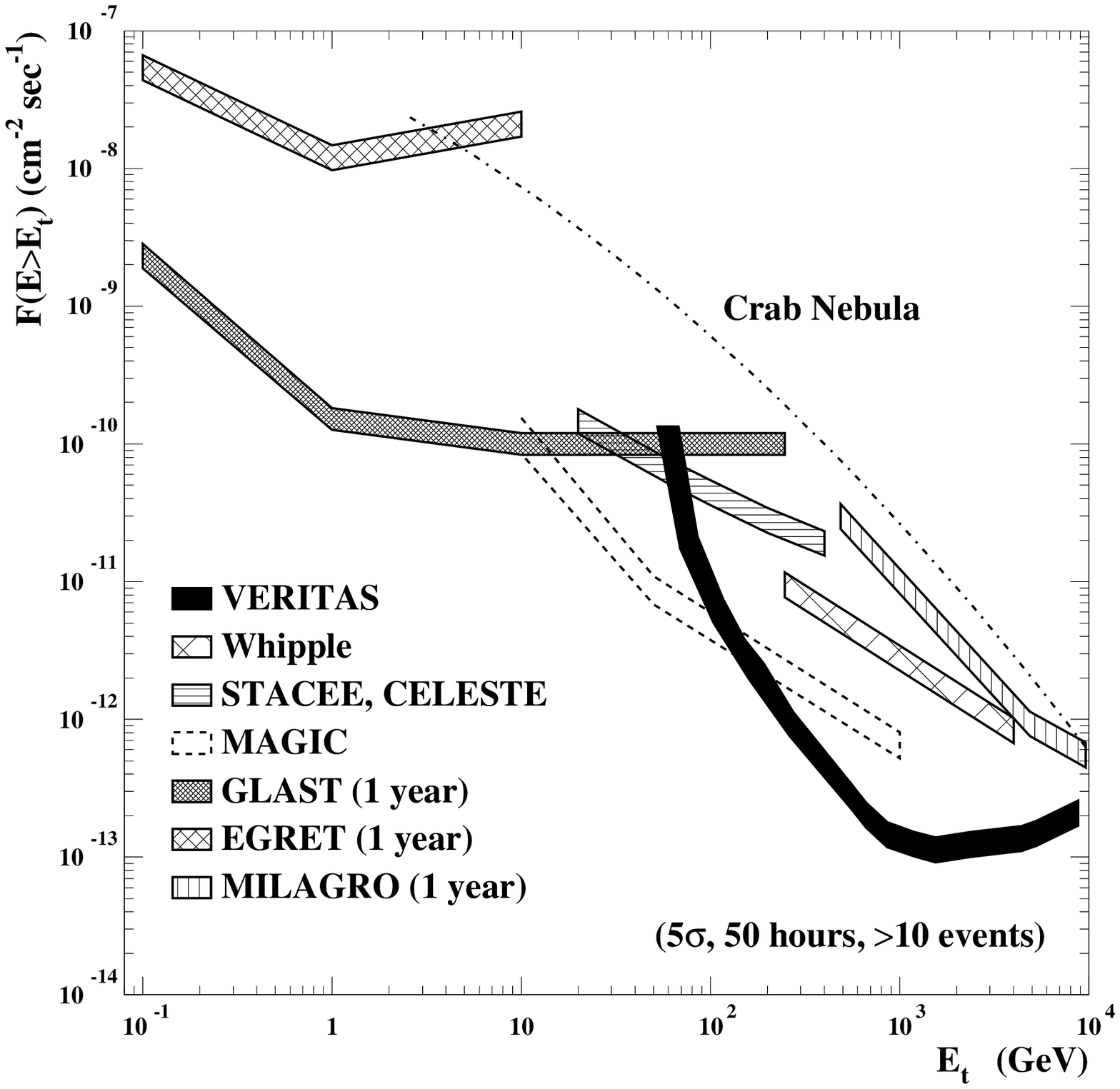} &
\includegraphics[width=6cm]{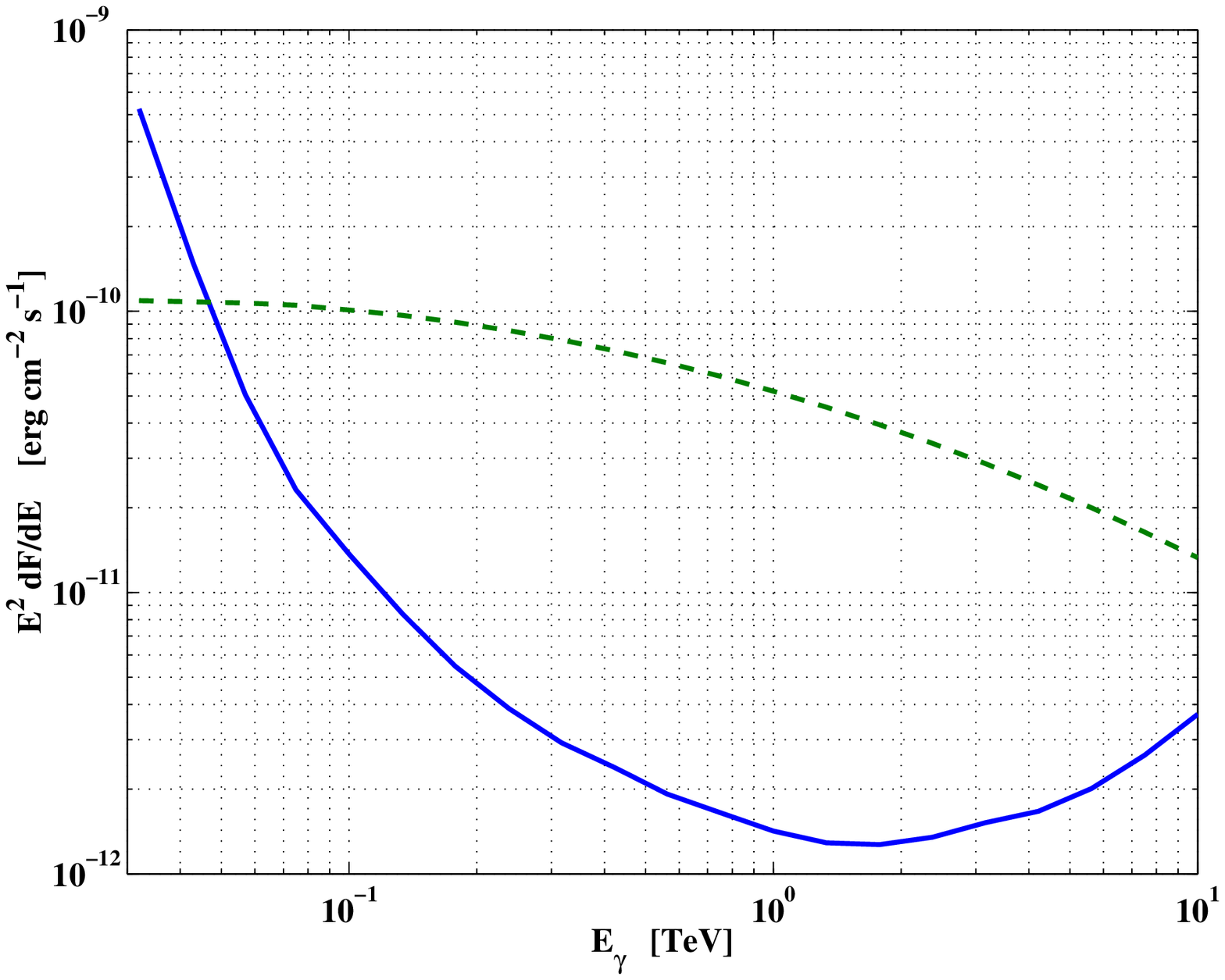} 
\end{tabular}
}
\caption{a. The integral flux sensitivity of past and future detectors
assuming 50 hour exposures for Whipple, HESS/VERITAS, MAGIC,
STACEE/CELESTE and one year of operation for Milagro, EGRET and GLAST.
b. Differential sensitivity of VERITAS for
50 hour exposure where the
source is detected to the 5$\sigma$ level per quarter decade of energy; 
this is the conservative sensitivity level where meaningful physical 
measurements can be made. 
The flux from the Crab Nebula is shown as a dashed line.}
\label{sensitivity}
\end{figure}

The integral flux sensitivities of a variety of techniques, both those 
currently in use and those that are under construction, are summarized in 
Figure~\ref{sensitivity}a; the differential sensitivity for VERITAS, 
a next generation telescope, is shown in Figure 2.b.
 
\section{A New Astronomical Window}

\begin{figure}
\centerline{\includegraphics[width=13cm]{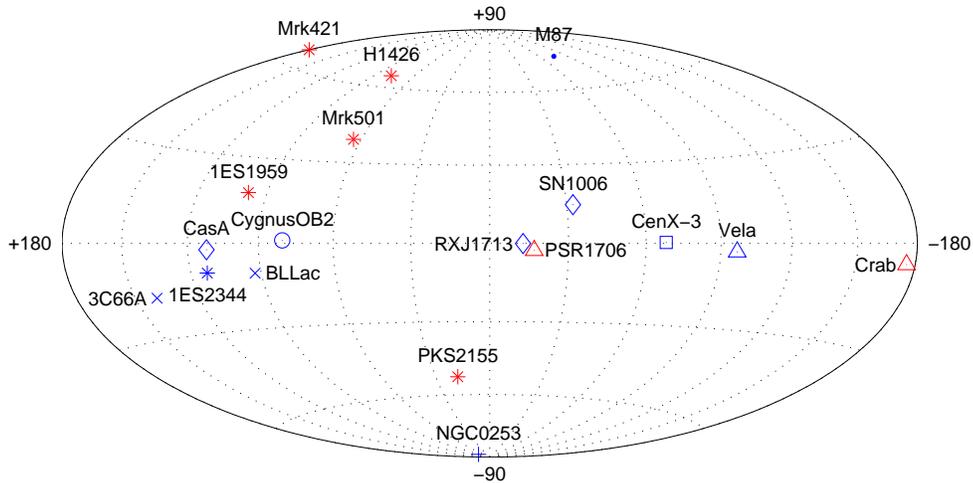}}
\caption{The distribution of known TeV sources in galactic coordinates [15].
Key: star=HBL; x=LBL; dot=radiogalaxy; +=starburst galaxy; 
diamond=SNR; triangle=plerion; square=binary; o=OB Association}
\label{map}
\end{figure}

Extensive observations over the past decade have led to the detection
of both Galactic and extragalactic sources.
The current catalog of sources is shown in Table~\ref{catalog}
[15] and plotted in Figure~\ref{map}  The criteria for inclusion
in this catalog are that the results should have been statistically significant
and have been published in a refereed journal.
It is noteworthy that many of these sources
are not in the EGRET Catalog [13], an indication
that the TeV sky opens a new window on the universe. The allotted
grade gives some measure of the credibility that should be assigned
to the reported detections; ``A'' sources have been independently
verified at the 5 $\sigma$ level whereas ``C'' sources clearly require
confirmation.  The fact that many of the
early entries to this catalog (based on single telescope observation)
have increased in significance with time suggests that systematic
effects in most experiments are understood and accounted for.  
Not included in the catalog are the several sources reported by the
CANGAROO-III group at the 28th ICRC in Tsukuba, Japan in August, 2003
[21] because they have not yet gone through the refereeing process.
These included the supernova remnants, RXJ0852.0-4622, GC40.5-0.5 and RCW86,
as well as the Galactic Center. A possible detection of the Galactic plane
was also reported by the Milagro group. The HEGRA/HESS groups reported
the confirmation of the blazars PKS2155-304 ($>$ 10$\sigma$),
1ES2344+514 (4.4$\sigma$), and BL Lac (3.0$\sigma$); these new results are 
reflected in the grade assigned to the sources.

\begin{table}[ht!]
\caption{Source Catalog c.2003 [15].}
\label{catalog}
\begin{center}
\begin{tabular}{llllll} \hline  

Catalog Name & Source &   Type &  Discovery & EGRET  & Grade  \\
       &        &   &   Date/Group & 3rd. Cat. & \\
\hline
\\
       
TeV 0047$-$2518 & NGC 253 & Starburst & 2003/CANG. & no & B \\
TeV 0219+4248 & 3C66A & Blazar & 1998/Crimea & yes & C$-$\\
TeV 0535+2200 & Crab Nebula & SNR & 1989/Whipple & yes & A \\
TeV 0834$-$4500 & Vela & SNR & 1997/CANG. & no & C \\
TeV 1121$-$6037 & Cen X-3  & Binary & 1999/Durham &  yes &  C \\
TeV 1104+3813 & Mrk 421 & Blazar & 1992/Whipple &  yes &  A \\
TeV 1231+1224 & M87 & Radio Gal. & 2003/HEGRA & no & C \\
TeV 1429+4240 & H1426+428 & Blazar & 2002/Whipple & no & A \\
TeV 1503$-$4157 & SN1006 & SNR & 1997/CANG. & no & B \\
TeV 1654+3946 & Mrk 501 & Blazar & 1995/Whipple & no &  A \\
TeV 1710$-$2229 & PSR 1706-44 & SNR & 1995/CANG. & yes & A \\
TeV 1712$-$3932 & RXJ1713.7-3946 & SNR & 1999/CANG. & no & B+ \\
TeV 2000+6509 & 1ES1959+650 & Blazar & 1999/TA & no & A \\
TeV 2032+4131 & CygOB2 & OB assoc. & 2002/HEGRA & yes? & B \\
TeV 2159$-$3014 & PKS2155-304 & Blazar & 1999/Durham & yes &  A \\
TeV 2203+4217 & BL Lac & Blazar & 2001/Crimea & yes & C \\
TeV 2323+5849 & Cas A  & SNR & 1999/HEGRA & no & B \\
TeV 2347+5142 & 1ES2344+514 & Blazar & 1997/Whipple & no & A \\

\hline
\end{tabular}
\end{center}
\end{table}

With an ever growing list of sources in the TeV catalog, it is
clear that the discipline has now reached some level of maturity.
Sources are no longer just detected; their spectra are measured and
their time variability characterized. Correllations are made over
many bands of the electromagnetic spectrum and there is little
doubt about the reality of most of the sources. In short, 
atmospheric Cherenkov experiments have left the domain of a
subsection of OG cosmic ray physics and become a legitimate 
branch of astronomical research. Its vocabulary is increasingly
that of astronomy and probably more alien to cosmic ray ears. 
Important new
results are presented at astronomical conferences and the frequency
of international workshops and symposia devoted to the field is
more than one per year. 
 
\section{The Origin of Cosmic Rays}

One of the chief motivations for the early efforts in gamma-ray
astronomy was that the detection of gamma-ray sources might solve
the mystery of the origin of cosmic rays. The scenario for the
solution was straightforward and perhaps simplistic. A source of
gamma rays with energy spectrum covering many decades would be
found. It would be apparent from the characteristic gamma-ray
spectrum that the gamma rays were produced by
the decay of $\pi^\circ$'s produced in the collision of cosmic-ray
protons with nuclear matter, e.g. hydrogen; 
a hard spectrum would be measured that mirrored that of the cosmic-ray 
protons with a peak in the spectrum at the 70 MeV.
If the cosmic radiation was to come from this source (or similar
ones), then since passage through interstellar space would soften
the exponent of the spectrum by about 0.5, the injection spectrum
index would be $\approx$ proportional to -2.0 to -2.2. This source, or
class of sources, would have to be sufficiently strong to satisfy
the overall power requirement of the observed cosmic-ray density.
The detection of a single source with the required spectrum to
indicate the presence of cosmic ray protons is interesting, of
course, but does not necessarily prove that this source of class of
source is THE source of the cosmic radiation.

Early estimates of anticipated fluxes were optimistic and
the experiments were more difficult than expected. Hence by the time
that reliable detections of sources were being reported, theory had
bypassed observation and the canonical view was that there really
was no cosmic ray mystery: cosmic rays, at least up to energies of
100 TeV, originated in supernova remnants in the Galaxy. Elegant
theories of the acceleration of particles in supernova shocks
supported this hypothesis whose only weak point was that there was
no direct experimental evidence to support them! Refined models
predicted there would be detectable fluxes of TeV gamma rays from
nearby remnants that had reached their Sedov phase [5]. 
The detection of a peak in the gamma-ray
spectrum near 70 MeV and a flat power law spectrum out to energies
of 100 TeV would be strong supporting evidence for this model. The
hypothesis received support from the apparent detection of several
supernova remnants by EGRET [7] although the 70 MeV peak was
not seen. Failure to see a gamma-ray spectrum from these sources
extending to TeV energies [3] seemed to contradict the
hypothesis that these sources constituted the origin of the cosmic 
radiation. Subsequent investigations (e.g. [10]) showed that
the presence of a cosmic electron component within a shell SNR
greatly complicated the interpretation of the gamma-ray spectrum
and suggested that even TeV observations would not be unambiguous
indicators of cosmic-ray acceleration. This has been demonstrated
for a number of individual sources where both hadron and electron
progenitor explanations have been advanced. 

In summary, we should still keep an open mind on the origin of 
the cosmic radiation. As noted in a recent report of a working group
composed of some of the leading theorists in the field
[6], ``Contrary to the general `folklore', it is by no 
means certain that SNRs are the source of the GCR and in fact the 
existence of the `knee' and the particles above the `knee'
is fairly clear proof that something else is required''.

The surprise of TeV gamma-ray astronomy has been not only the 
number of sources detected but the wide variety of source 
classes (Table~\ref{catalog}). No less than six 
different types of source are seen:
supernova remnants, a binary, an OB association, blazars, a radio galaxy 
and a starburst galaxy. This suggests that the acceleration of high 
energy particles is a common phenomenon on both Galactic and 
extragalactic scales. What is really surprising (apart from the number and 
diversity of TeV sources) is that in no instance is there a 
completely unambiguous identification of a source of hadrons.  
Although all cosmic ray physicists yearn for an explanation of
cosmic ray origins, it would be rash to assume that TeV astronomy
has yet provided the smoking gun to solve this mystery. It may be that
the observations of
our signal-poor, but resource-rich, cousins in TeV neutrino
astronomy will be required to provide this smoking gun.

\section{Extragalactic Sources}

Given that there is a consensus in the cosmic-ray community 
that cosmic rays at energies below 100 TeV are a Galactic phenomenon
it comes as a surprise that there are more extragalactic than Galactic 
sources in the catalog (Table~\ref{catalog}). This is particularly 
apparent if the catalog is restricted to grade ``A'' sources. It is
somewhat surprising that there should be a surfeit of extraglactic 
sources when the assumed origin of the observed cosmic radiation 
at these energies is Galactic (albeit the detections are aided by 
relativistic beaming!).

\subsection{New Sources}

The most interesting new results in TeV extragalactic astronomy are 
undoubtably the reported detections of the starburst galaxy, NGC253 
and the radio galaxy, M87. These detections are sufficiently 
new that they have not yet been confirmed but they open new possibilities
for the sources of the extragalactic cosmic radiation.

{\it NGC 253: }
This is the first starburst galaxy detected and also the closest
(250 kpc). Starburst galaxies are the site of extraordinary
supernovae activity and were postulated to be sources of VHE cosmic
rays and gamma rays [25]. The reported detection by
CANGAROO-II in 2002 was at the 11 $\sigma$ level [16]. It
was observed to have a very steep spectral index (-3.75) which
implies that most of the signal is close to the telescope
threshold and therefore more subject to systematic effects. 
The TeV source was extended with the same elongation as the
optical source. 

{\it M87:}
This is one of the brightest nearby radio galaxies and is an
obvious potential source of high energy radiation since the jet
displays evidence for synchrotron radiation and time variability.
The angle of the jet is about 30$^\circ$ which means that it is
unlikely to have the same observational gamma-ray properties as
the blazars. In fact it is not observed by EGRET and the positive
observation by the HEGRA group [1] was a surprise.
Although the detection was only at the 4.7 $\sigma$ level of
significance (weaker than any of the other sources in the TeV
catalog) it is potentially exciting result as it opens up the
possibility that many AGN may be observable whose axes are not
pointing directly towards  us.
It is, at best, a weak source and its detection
required 83 hours of observation. It was not seen in observations
at lower energies [20] but the
exposures, and hence the flux sensitivities were limited. The
detection of M87 revives interest in the reported detection of
Centaurus A in 1975 [11] which, although not
confirmed in later, more sensitive, observations, was at a time
when the source had an abnormally high microwave flux.

\subsection{Blazars}

The clearest evidence that TeV gamma-ray astronomy is a mature
scientific discipline comes from consideration of the catalog of 
blazars detected at TeV energies [15]; this is 
summarized in Table~\ref{blazars} There is a wealth of data
about the properties of this class of sources. Some of these are
summarized below but more complete descriptions are 
given elsewhere [15, 27]. This rich
database, which includes source positions, distances,
fluxes, spectra, time variability and multi-wavelength
correlations surely demonstrates that this is a viable 
new branch of astronomy.

\begin{table}
\caption{Blazar Catalog of TeV Gamma-ray Sources}
\label{blazars}
\begin{center}
\begin{tabular}{|l|l|l|c|} \hline  
Catalog Name & Source & Classification & Redshift\\
\hline
TeV 0219+4248 & 3C66A & BL Lac (LBL) & 0.444 \\
TeV 1104+3813 & Mrk 421 & BL Lac (HBL)  & 0.031\\
TeV 1429+4240 & H1426+428 & BL Lac (HBL) & 0.129\\
TeV 1654+3946 & Mrk 501   & BL Lac (HBL) & 0.033\\
TeV 2000+6509 & 1ES1959+650 & BL Lac (HBL) & 0.048\\
TeV 2159$-$3014 & PKS2155$-$304 & BL Lac (HBL) & 0.116\\
TeV 2203+4217 & BL Lac & BL Lac (LBL) & 0.069\\
TeV 2347+5142 & 1ES2344+514 & BL Lac (HBL) & 0.044 \\
\hline
\end{tabular}
\end{center}
\end{table}

\begin{figure}[ht!]
\centerline{%
\begin{tabular}{c@{\hspace{6pc}}c}
\includegraphics[width=6cm]{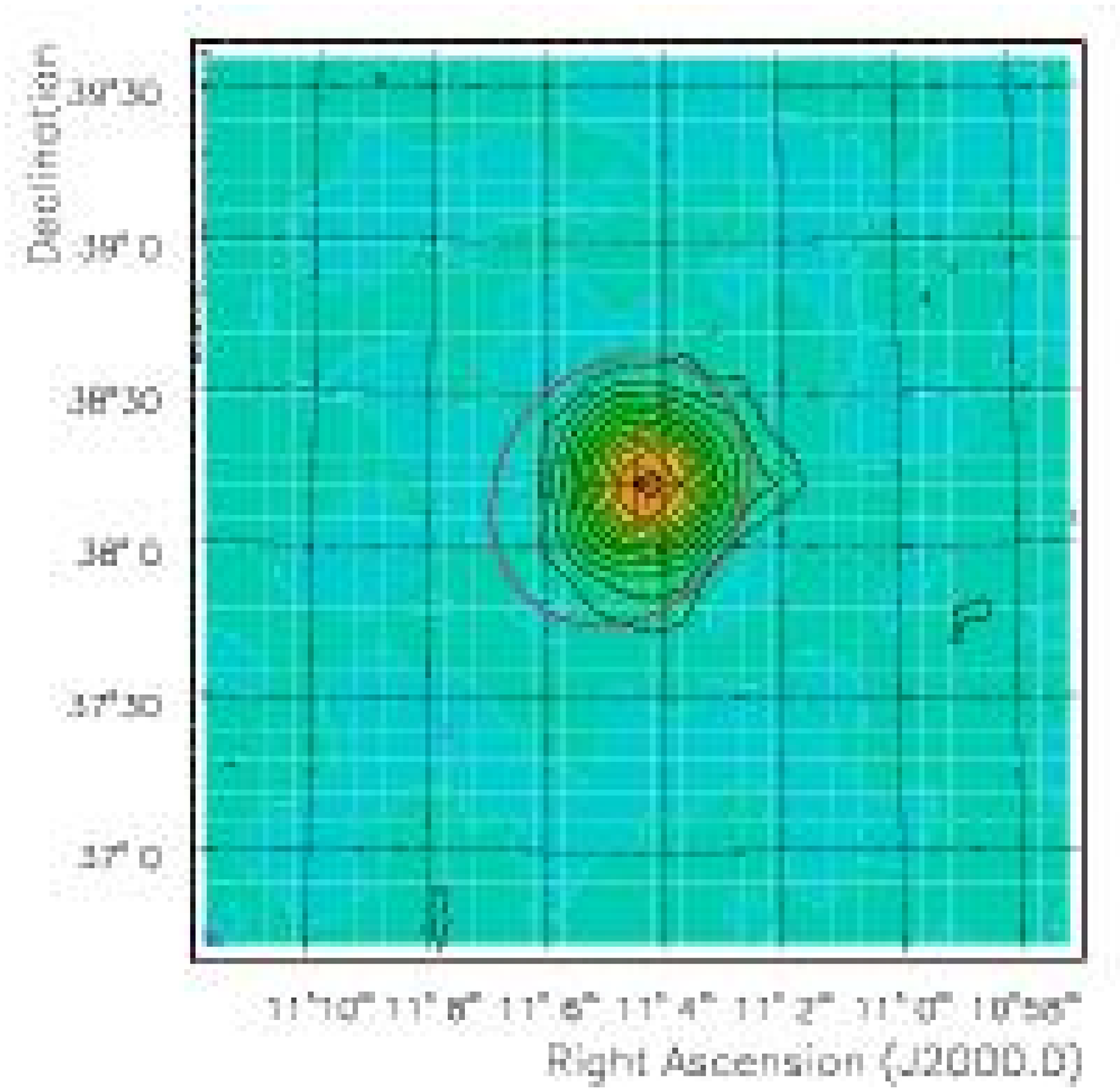} &
\includegraphics[width=8cm]{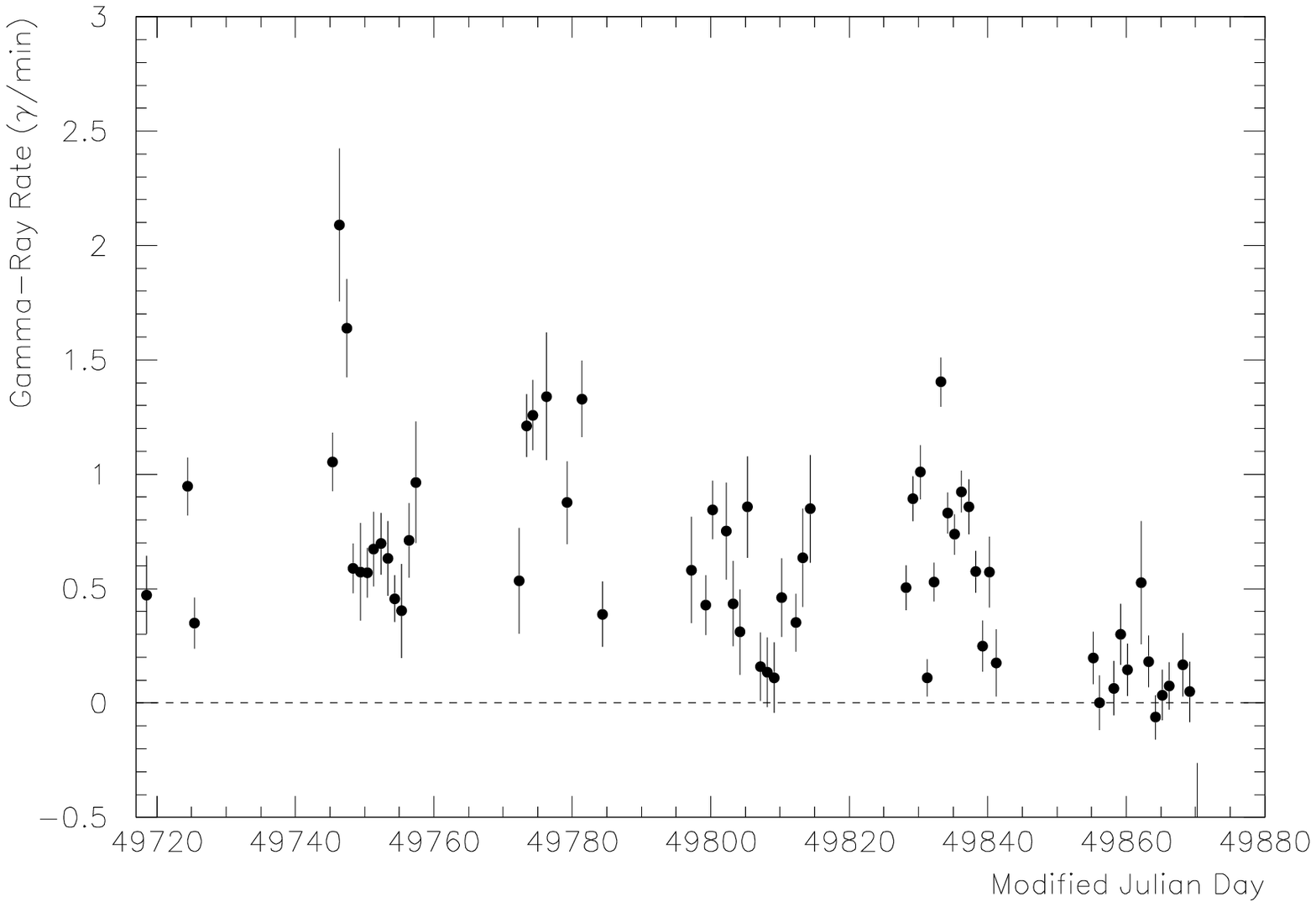}
\end{tabular}
}
\caption{a. Location of Mrk 421; b. Variability on nightly scale 
from Mrk 421}
\label{location}
\end{figure}

The ability of the new discipline to fix the locations and 
dimensions is illustrated by observations of Mrk\,421.
This was detected as
the first extragalactic source of VHE gamma rays in 1992 by the
Whipple Observatory gamma-ray telescope.  A
two-dimensional image of Mrk\,421 in TeV gamma rays is shown in
Figure~\ref{location}a.  The
uncertainty in the source location at VHE energies
(0.05$^\circ$) was significantly less than at HE energies
(0.5$^\circ$)
because Mrk 421 is such a weak source at 100 MeV energies. 
This detection illustrates the ability of TeV telescopes to 
fix the location of a strong source (a few hundred photons).
Unlike some of the Galactic TeV sources none of the blazars shows 
any evidence for being extended, at least to the present 
resolution of a few arc-min.

All of the VHE blazars detected to date are
relatively close-by with redshifts ranging from 0.031 to 0.129, and
perhaps to 0.44. The well-established TeV blazars
(A sources) are High Frequency BL Lacs (HBLs). These
are sources whose synchrotron peaks occur near the X-ray band.
Such blazars seem more likely to be sources of TeV gamma rays than LBLs 
since the presence of higher energy electrons is implied.

Extreme variability on time-scales from minutes to
years is the most distinctive feature of the VHE emission from
blazars. Clear evidence for flaring activity
in the TeV emission of Mrk\,421 in 1995 is  shown in
Figure~\ref{location}b.

\begin{figure}[ht!]
\centerline{%
\begin{tabular}{c@{\hspace{6pc}}c}
\includegraphics[width=6cm]{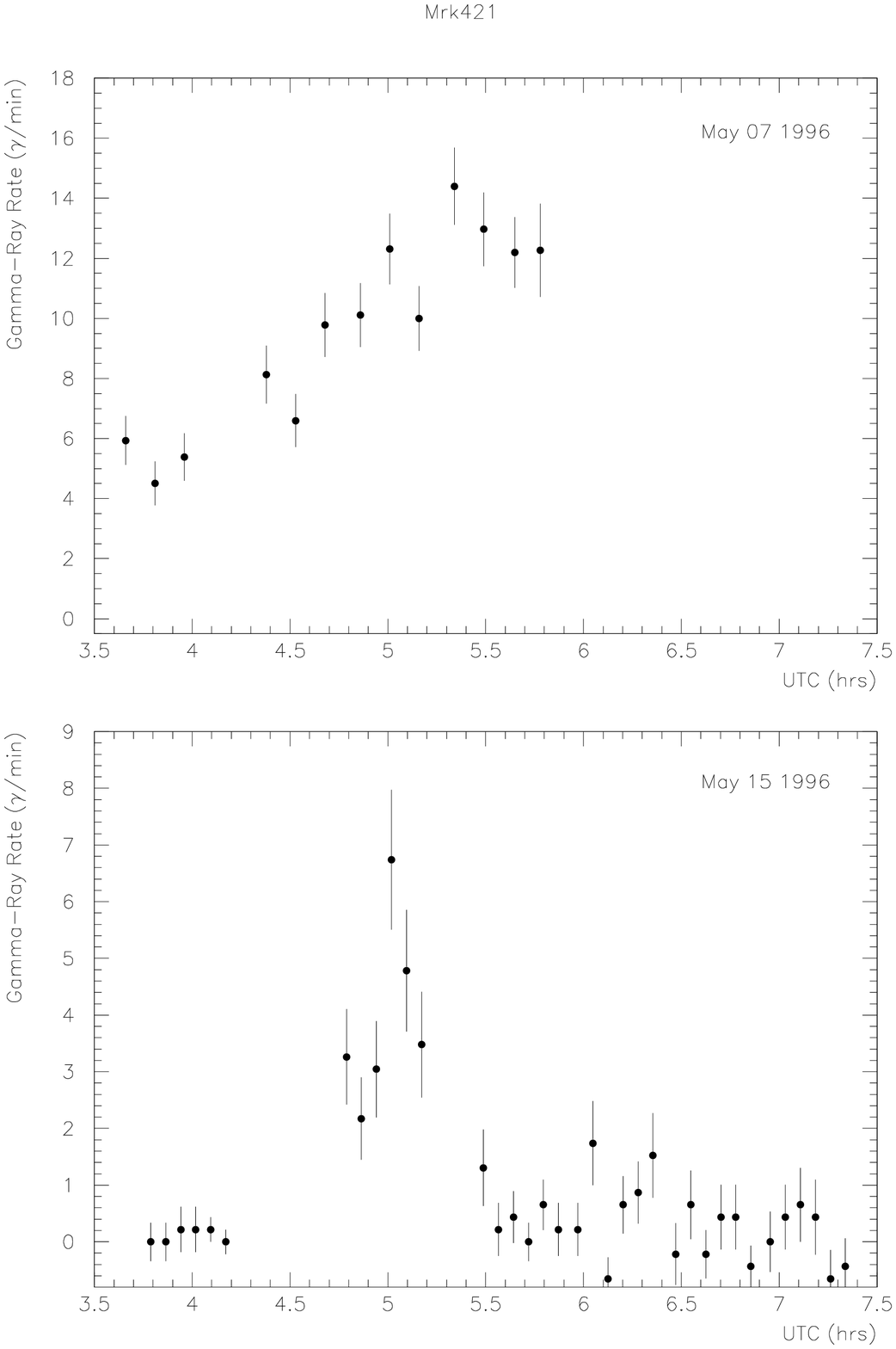} &
\includegraphics[width=6cm]{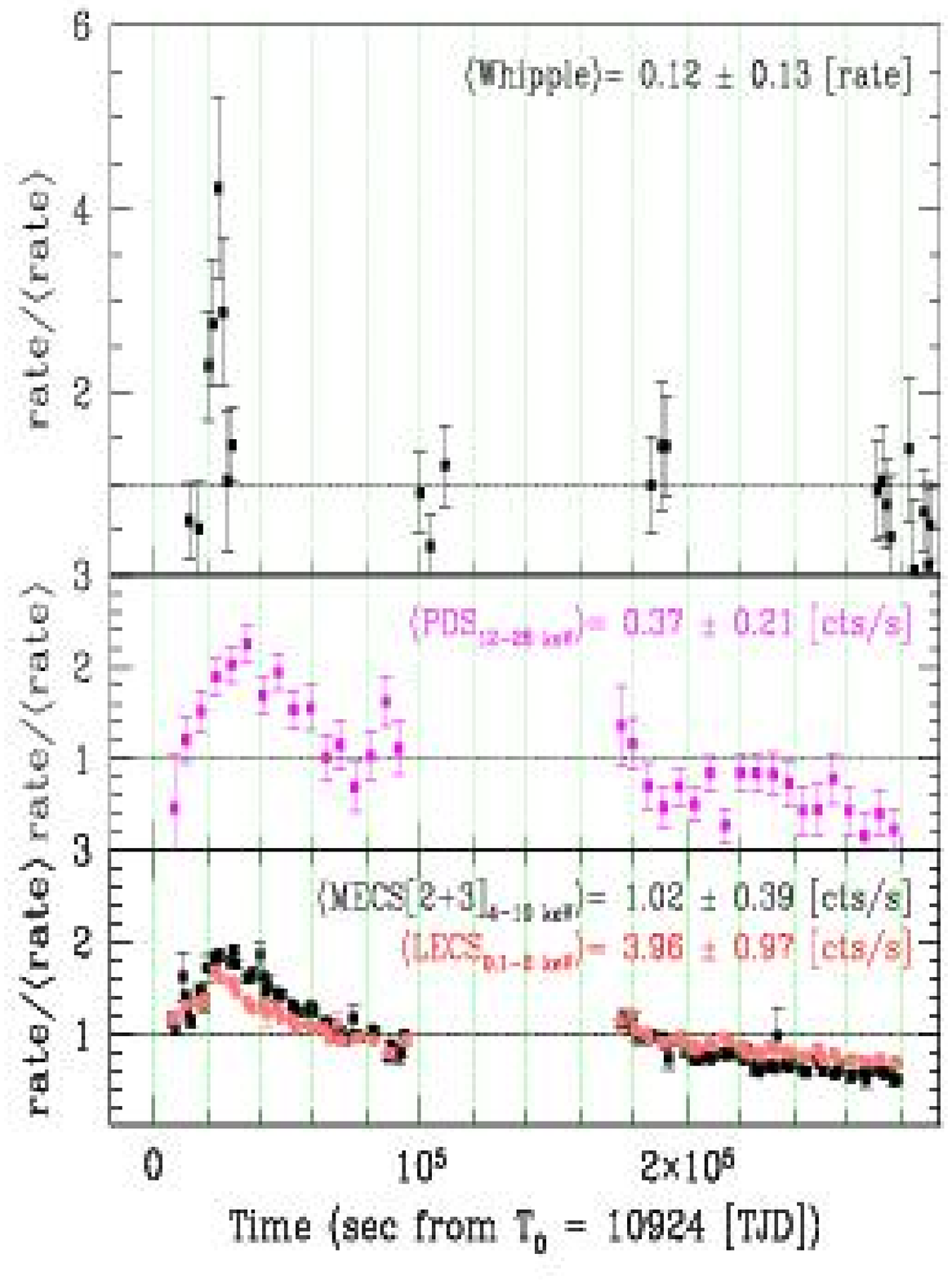}
\end{tabular}
}
\caption{a. Sub-hour variations in Mrk 421 as seen at the Whipple Observatory; 
 b. X-ray and gamma-ray variations in Mrk 421.}
\label{variations}
\end{figure}

The observations of two short flares [9]
(Figure~\ref{variations}a)
were a dramatic demonstration of the rich time structures that
are present in blazars.  In the first flare,
observed on May 7, 1996 the flux increased monotonically during the
course
of $\sim$2 hours of observations. This flux is the highest observed
from any VHE
source to date.  The doubling time of the flare was $\sim$1 hour.
A second flare, observed on May
15, 1996, although weaker, was remarkable for its very short duration: the
entire flare lasted approximately 30 minutes with a doubling and
decay time of less than 15 minutes.  These two flares exhibited the
fastest time-scale variability, by far, seen from any blazar at any
gamma-ray energy.

\begin{figure}[ht!]
\centerline{%
\begin{tabular}{c@{\hspace{6pc}}c}
\includegraphics[width=8cm]{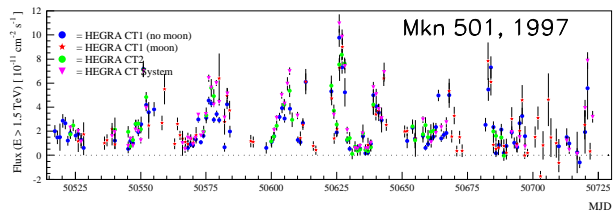}
\end{tabular}
}
\caption{Variations in Mrk 421 recorded by HEGRA over a six month period.}
\label{hegra}
\end{figure}

Variations are also seen on much slower time-scales as well.
In 1997, the VHE emission from Mrk\,501 increased dramatically.
After being the weakest known source in the VHE sky in 1995-96, it
became the brightest; the amount of day-scale flaring
increased and, for the first time, significant hour-scale
variations were seen. The six month history of observations by the
HEGRA telescope [17] is shown in Figure~\ref{hegra}b.

\begin{figure}[ht!]
\centerline{%
\begin{tabular}{c@{\hspace{6pc}}c}
\includegraphics[width=6cm]{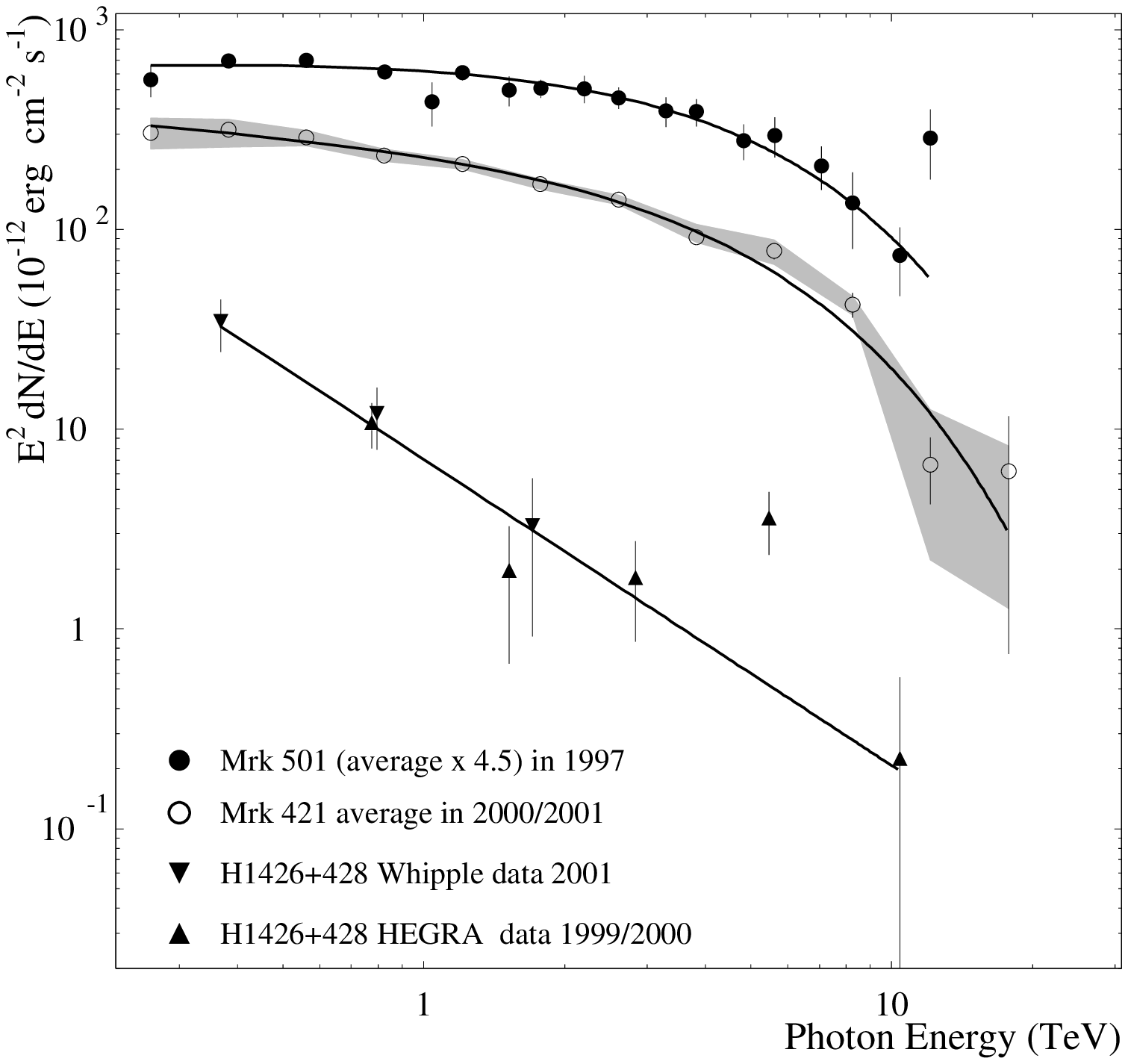} &
\includegraphics[width=7cm]{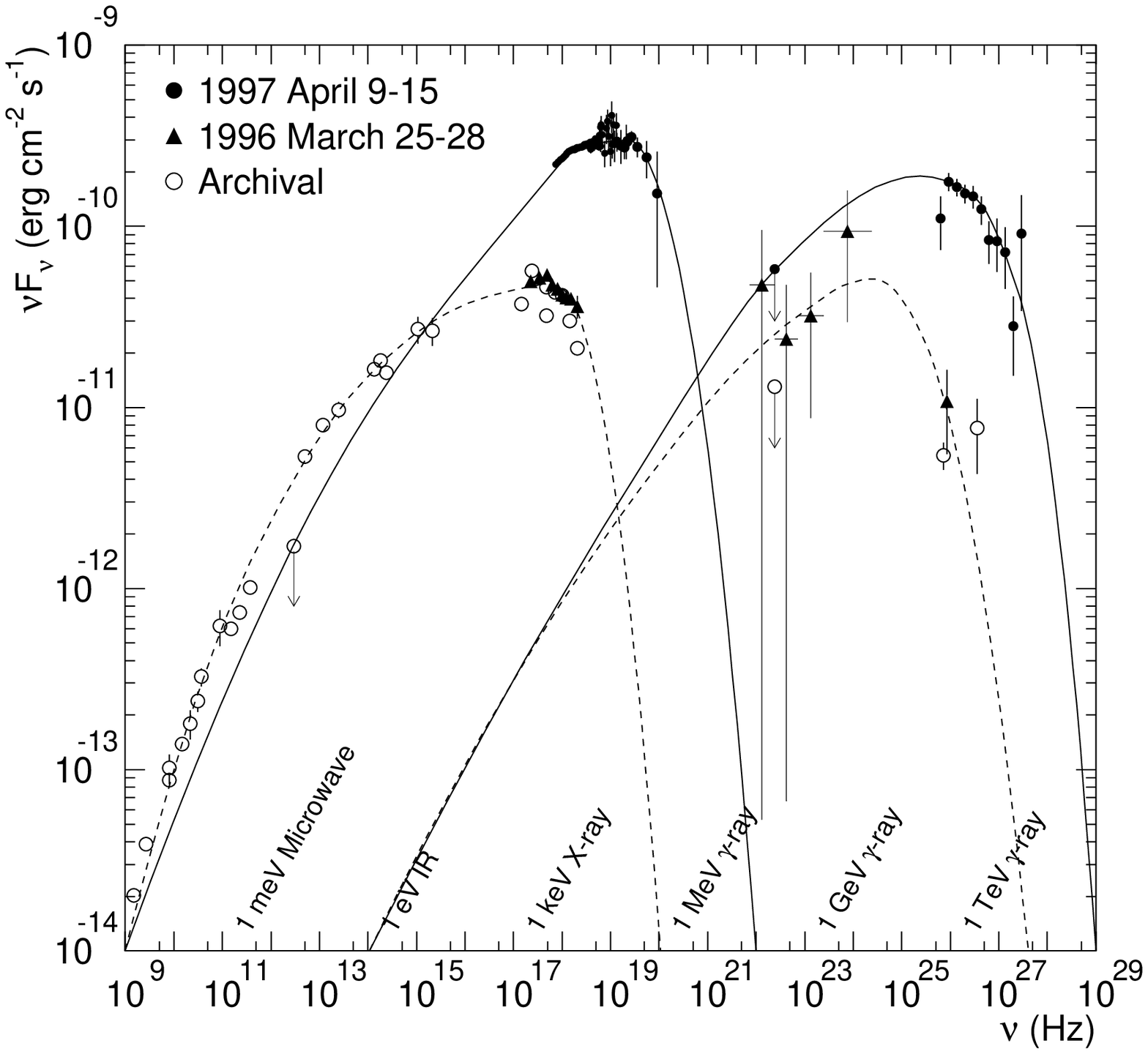}
\end{tabular}
}
\caption{a. Spectral measurements of Mrk, 421, Mrk 501 and H1426+42 [19]; 
b. SED of Mrk 501}
\label{SED}
\end{figure}

The high flux VHE emission from Mrk\,501 in
1997 and Mrk\,421 in 2001 has permitted detailed spectra to be
extracted. Measurements
were possible over nearly two decades of energy. As many as 25,000
photons were detected in these outbursts so that the spectra were
derived with high statistical accuracy. Unlike the HE sources where
the photon-limited blazar measurements are consistent with a simple
power law, there is definite structure seen in the VHE
measurements. The spectra of Mrk 421 and Mrk 501 can each be fit
with a spectrum with an exponential cutoff (Figure~\ref{SED}a).

For Mrk 421, the exponential cut-off energy is $\approx$ 4 TeV and
for Mrk 501 it is $\approx$ 3-6 TeV. The coincidence of these two values
suggests a common origin, i.e., a cut-off in the acceleration
mechanisms within the blazars or perhaps the effect of the infra-red
absorption in extragalactic space. Attenuation of the VHE gamma
rays by pair-production with background infra-red photons could
produce a cut-off that is approximately exponential. The TeV signal
from Mrk 421 in 2002
was sufficiently strong that the data could be divided into hourly intervals and
spectrally analyzed; the results show that the spectrum clearly
hardens with total intensity but the same exponential cut-off can be
fitted to all the data [18, 19].

Because of the relative flexibility of 
observations with ground-based telescopes,
it has been possible to organize some extensive multi-wavelength 
campaigns so that
the Spectral Energy Distributions (SEDs) of the TeV blazars are 
much better determined than those
of the EGRET blazars.  Observations of Mrk 421
at TeV energies with the Whipple telescope and at X-ray wavelengths
with the {\it BeppoSAX} satellite, established the first hour-scale
correlations between X-rays and gamma rays in a blazar.  The
light-curve for the observations by {\it BeppoSAX} in three X-ray bands
and Whipple above 2\,TeV is shown in Figure~\ref{variations}b.

Figure~\ref{SED}b shows the SEDs
expressed as power per logarithmic bandwidth, for Mrk\,421 and
Mrk\,501 derived from contemporaneous multi-wavelength observations
and an average of non-contemporaneous archival measurements.  Both
have a peak in the synchrotron emission at X-ray frequencies, which is
typical of HBLs, and a high energy peak whose
exact location is unknown but must lie in the 10 -- 250\,GeV range.
Both the synchrotron and high energy peak are similar in power
output, unlike the EGRET-detected flat spectrum radio sources which can
have their ``Compton'' power peaks well above the synchrotron power peaks.  

\section{Prospects}

\begin{figure}[ht!]
\centerline{%
\begin{tabular}{c@{\hspace{6pc}}c}
\includegraphics[width=6cm]{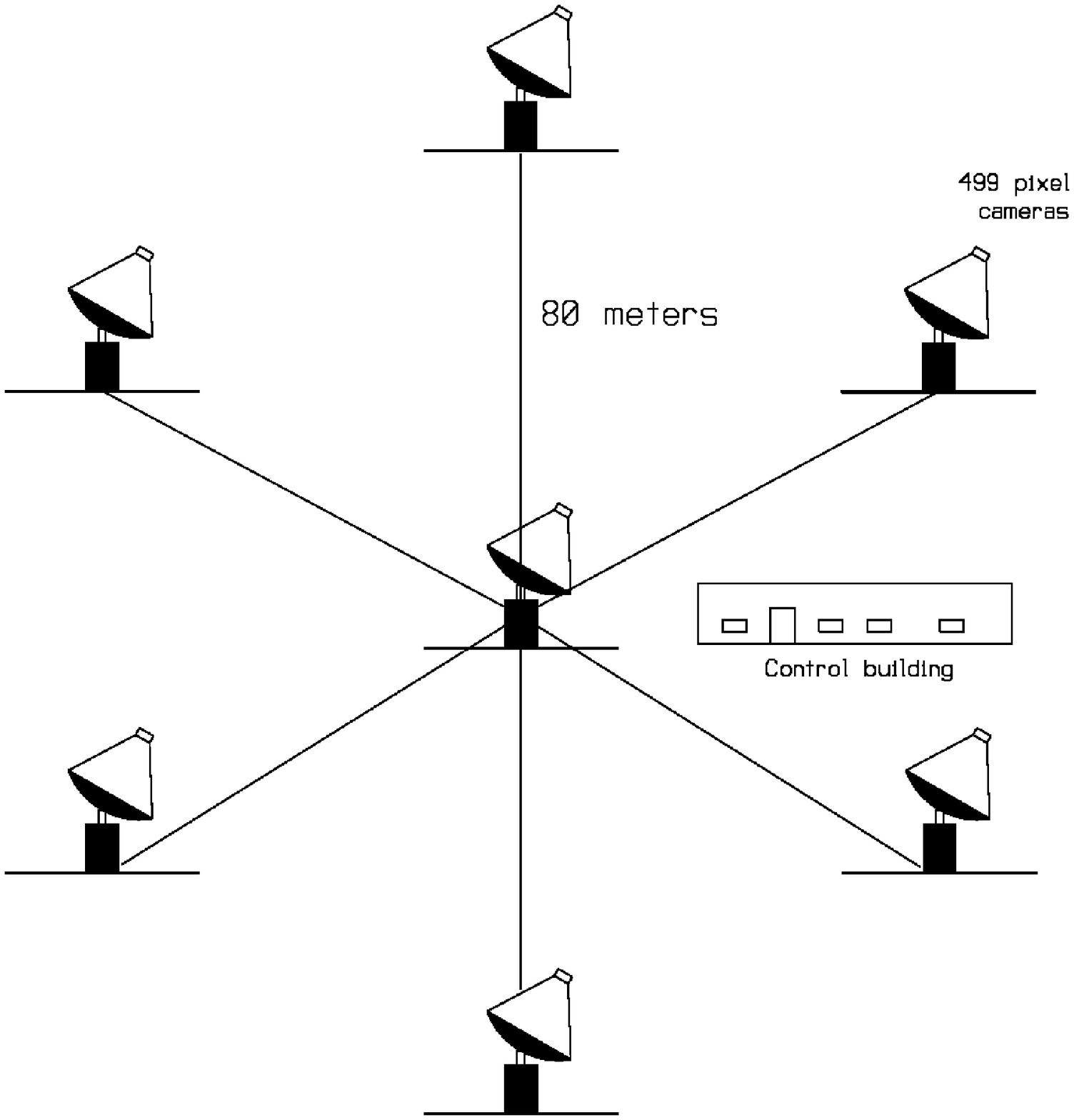} &
\includegraphics[width=6cm]{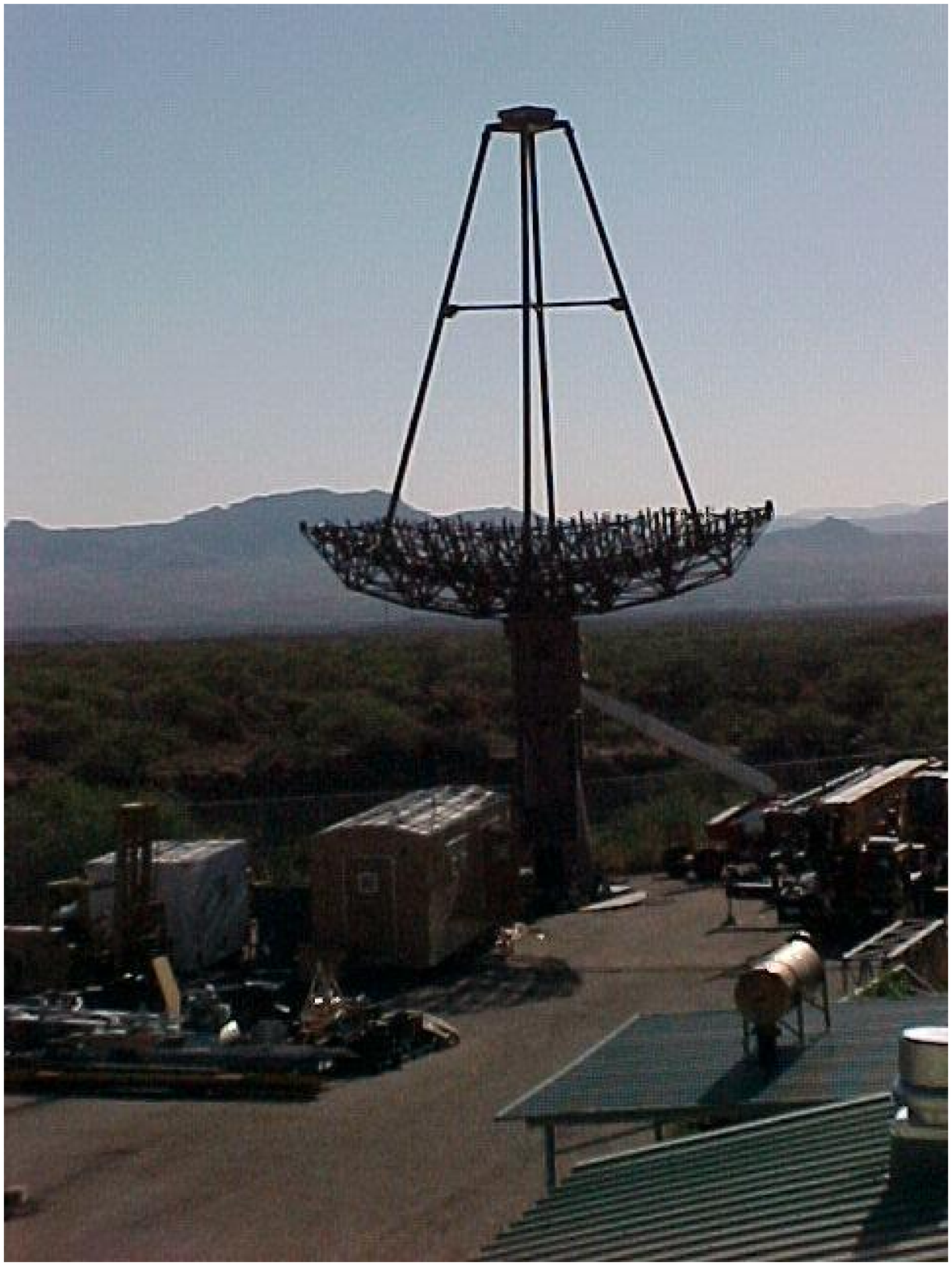}
\end{tabular}
}
\caption{(a) The VERITAS array layout.
(b) VERITAS Prototype Telescope.}
\label{veritas}
\end{figure}

The third-generation systems (CANGARO0-III, HESS, MAGIC, VERITAS) that
are now under construction will dominate the VHE observational arena 
for the next decade. The Very
Energetic Radiation Imaging Telescope Array System (VERITAS) was
the first of these next generation telescopes to be proposed but
will probably be the last to be built. The
seven, 12 m aperture, telescopes in VERITAS will be identical and 
will have the
geometrical layout shown in Figure~\ref{veritas} Six telescopes
will be located at the corners of a hexagon of side 80\,m, and one
will
be located at the center. The telescopes will each have a camera
consisting of 499 pixels with a field of view of 3.5$^{\circ}$
diameter.
A feature of this array will be the flexibilty offered with the
possiblity of operating with the telescopes in different configurations.

The most exciting aspect of the recent VHE results is the diversity of
objects that are now proving to be VHE gamma-ray sources; many of
them have not been detected by EGRET. With the improved sensitivity
of GLAST, hopefully the 100 MeV component of these sources
will also be detectable.
Although the detection of unidentified TeV sources is still in
its infancy, the revelation that the TeV sky map is quite different
from that at 100 MeV suggests that there may be many surprises in
store.

\section{Acknowledgments}

This research is supported by a grant from the U.S. Department of
Energy. I am grateful to Deirdre Horan and Frank Krennrich for their 
comments on the paper.

\section{References}

\vspace{\baselineskip}

\re
1.\ Aharonian, F.A. et al., \ 2003, A \& A, 403, L1; astro-ph/0310308
\re
2.\ Bhat, C.L., \ 1998, Proc. 25th I.C.R.C., (1997, Durban),
Publ. World Scientific, Eds. M.S.Potgieter et al., 211
\re
3.\ Buckley, J.H. et al., \ 1998, A \& A, 329, 639
\re
4.\ Buckley, J.H, \ 2000,  26th I.C.R.C. (1999, Salt Lake 
City), AIP Conf. Proc. 516, Eds. B.Dingus et al., 195 
\re
5.\ Drury, L.O'C., \ et al. \ 1994, A \& A, 287, 959
\re
6.\ Drury, L.O'C., \ et al. \ 2002, Space Sci. Rev., 99, 329
\re
7.\ Esposito, J.A., \ et al. \ 1996, ApJ, 461, 820
\re
8.\ Fegan, D.J., \ 1990, Proc. 21st I.C.R.C. (1990, Adelaide),
Publ. Univ. of Adelaide, Ed. R.J.Protheroe, 11, 23
\re
9.\ Gaidos, J.A. et al., \ 1996, Nature, 383, 319
\re
10.\ Gaisser, T.K. et al., \ 1998, ApJ, 492, 219  
\re
11.\ Grindlay, J.E. et al., \ 1975, ApJL, 197, L9 
\re
12.\ Haines, T.J., \ 1996, Proc. 23rd I.C.R.C. (1993, Calgary),
Publ. World Scientific, Eds. D.A.Leahy et al., 341
\re 
13.\ Hartman, R.C. et al. \ 1999, ApJ Suppl., 123, 79
\re
14.\ Hillas, A.M., \ 1996, Proc. 24th I.C.R.C. (1995, Rome), 
Il Nuovo Cimento, 19C, N5, 701
\re
15.\ Horan, D, Weekes, T.C., \ 2003, Proc. Symp. on ``TeV Astrophysics
of Extragalactic Sources'', (2003, Chicago), New Astronomy, in press 
\re
16.\ Itoh, C. et al. \ 2002, A\&A, 396, 379
\re
17.\ Kranich, D. et al., \ 1999, Proc. 26th I.C.R.C., (1999, 
Salt Lake City), 3, 358
\re 
18.\ Krennrich, F. et al., \ 2001, ApJL 560, L45; \ 2002 ApJL, 575, L9
\re
19.\ Krennrich, F. et., \ 2002, Symposium on ``The Universe Viewed in 
Gamma-Rays'', Kasiwa, Japan, September, 2002, Publ. Universal Academy,
Inc., 159
\re
20.\ Lebohec, S. et al., \ 2003, this conference
\re 
21.\ Mori, M., \ 2003, Proc. 28th I.C.R.C. (2003, Tsukuba),
Rapporteur, in press.
\re
22.\ Naito, T., Takahara, F., \ 1994, J.Phys.G.: Nucl.Part.Phys. 20, 477
\re 
23.\ Porter, N.A., \ 1983, Proc. 18th I.C.R.C., (1983, Bangalore), 
Publ. Tata Institute, Eds. N.Durgaprasad, 12, 435
\re
24.\ Protheroe, R.J., \ 1987, Proc. 20th I.C.R.C., (1987, Moscow),
8, 21
\re
25.\ Voelk, H.J. et al, \ 1996 Space. Sci. Rev. 75, 279 
\re
26.\ Weekes, T.C. \ 1991, Proc. 22nd I.C.R.C. (Dublin, 1991),
Publ. D.I.A.S., 5, 59
\re
27.\ Weekes, T.C., \ 2003, ``Very High Energy Gamma Ray Astronomy'',
Inst. of Phys. (U.K.).
\endofpaper
\end{document}